%% file: ms.tex
\documentclass[iop]{emulateapj} 
\usepackage{natbib,threeparttable,apjfonts,graphicx,booktabs}

\newcommand{\xmm}{\emph{XMM--Newton}}
\newcommand{\rosat}{\emph{ROSAT}}
\newcommand{\nH}{$N_{\text{H}}$}
\newcommand{\snr}{G296.1--0.5}
\newcommand{\ps}{2XMMi~J1150}
\newcommand{\mosi}{MOS~1}
\newcommand{\mosii}{MOS~2}

\shortauthors{Castro et al.}
\slugcomment{Accepted for publication in ApJ}

\begin{document} 
\title{An XMM-Newton Study of the bright, nearby supernova remnant G296.1--0.5}
\author{D. Castro\altaffilmark{1,2}, P. O. Slane\altaffilmark{1}, B. M. Gaensler \altaffilmark{3}, J. P. Hughes\altaffilmark{4}, and D. J. Patnaude\altaffilmark{1}}

\altaffiltext{1}{Harvard-Smithsonian Center for Astrophysics, 60 Garden Street, Cambridge, MA 02138, USA}
\altaffiltext{2}{Departamento de F\'isica, Universidad Sim\'{o}n Bol{\'\i}var, Valle de Sartenejas, Apdo. 89000, Caracas 1080A, Venezuela}
\altaffiltext{3}{Sydney Institute for Astronomy, School of Physics A29, The University of Sydney, NSW 2006, Australia}
\altaffiltext{4}{Department of Physics and Astronomy, Rutgers University, 136 Frelinghuysen Road, Piscataway, NJ 08854-8019}

\begin{abstract}
We present a detailed study of the supernova remnant \snr, performed using observations with the EPIC and RGS instruments of the \xmm\ satellite. \snr\ is a bright remnant that displays an incomplete multiple-shell morphology in both its radio and X-ray images. We use a set of observations towards \snr, from three distinct pointings of EPIC, in order to perform a thorough spatial and spectral analysis of this remnant, and hence determine what type of progenitor gave rise to the supernova explosion, and describe the evolutionary state of the SNR. Our \xmm\ observations establish that the spectral characteristics are consistent across the X-ray bright regions of the object, and are best described by a model of the emission from a nonequilibrium ionization collisional plasma. The study reveals that the emission from the shell is characterized by an excess of N and an underabundance of O, which is typical of wind material from red supergiant (RSG) and Wolf-Rayet (WR) stars. Additionally, we have detected transient X-ray source 2XMMi~J115004.8-622442 at the edge of the SNR whose properties suggest that it is the result of stellar flare, and we discuss its nature in more detail.  

\end{abstract}


\section{INTRODUCTION}

The supernova remnant (SNR) G296.1--0.5 is an impressively bright X-ray object that has also been studied at various other wavelengths. It  was first detected  by \citeauthor{Clark1973} (1973,1975), with the Molonglo Cross radio telescope, at a frequency of 408 MHz. In the radio band, the image of this object is characterized by a faintly defined shell of synchrotron emission, with angular size $\sim$33$'$. The radio emission is brightest towards a lobe in the southwest of the field, which in the early radio studies was the only emission attributed to this remnant.  A later study by \citet{Caswell1983} at the same frequency, suggested the structure of the remnant to be a somewhat elliptical ring, including emission from the northwest, a ridge of emission towards the east, and the bright lobe in the southwest quadrant. 

Optical studies revealed faint nebulosity in H$\alpha$ and [\ion{S}{2}], coincident with the northwest region of the radio remnant, with collisionally excited line spectra. 
These authors also observe strong [\ion{N}{2}] lines in the optical spectra of the filaments studied. \citet{Russeil2001} detected strong localized diffuse H$\alpha$ emission towards \snr\ using data from the Marseille H$\alpha$ survey, and thin filaments are evident in the AAO/UKST H$\alpha$ survey.

Observations with the {\it European X-Ray Observatory Satellite} ({\it EXOSAT}) and the {\it Einstein Observatory}'s IPC and HRI cameras \citep{Markert1981,Bignami1986} reveal a partial-shell morphology with three bright regions of X-ray emission. Although the X-ray image corresponds well with the radio map, the detailed features are poorly correlated, and there are considerable variations in the X-ray to radio brightness. The {\it Einstein} IPC count rate detected from this object is similar to those observed from well studied larger SNRs like W28, W44 and PKS 1209--51/52 \citep{Long1991,Rho1994,Matsui1988}, making it a remarkably bright X-ray source. 

A spectral study of G296.1--0.5 with the Position Sensitive Proportional Counter (PSPC) of the {\it R\"ontgen-Satellit} (\rosat) by \citet{Hwang1994}  suggested that the X-ray emission from this remnant is well-described by a single temperature plasma model in ionization equilibrium, with an overall temperature of $\sim$0.2 keV. This spectral fit requires extremely low abundances relative to solar ones, roughly 5\%, and substantial interstellar absorption ($N_{\text{H}} \gtrsim 1.5 \times 10^{21} {\rm\ cm}^{-2}$). An alternative plasma model, with two temperature components (approximately 0.1 keV and 0.3 keV) and solar abundances, was found to be in equally good agreement with the data. \citet{Hwang1994} use the optically estimated distance of 4 kpc, and an assumed Sedov expansion, to establish an approximate remnant age of 20,000 years, a swept-up interstellar medium (ISM) mass $\sim$250 M$_{\odot}$, and ambient density $\sim$0.8 cm$^{-3}$.

The low spectral resolution of \rosat, together with the poorly constrained models needed to describe the X-ray data, suggest that a deeper investigation of this remnant is required in order to determine its nature. In this paper, we present a study of new \xmm\ EPIC and RGS X-ray data of \snr, which aims to provide a more satisfying characterization of the X-ray emission from this object, and hence establish the evolutionary state of the remnant. The high-quality  spectral data of the study with \xmm\ is used to investigate spectral variations within the remnant and possible abundance enhancements, and provides a useful insight into the nature of \snr.  We describe the \xmm\ observations, and analysis of the results in Section 2. In Section 3 we discuss the nature of the X-ray emission, the properties of the remnant and the characteristics of its morphology, stage of evolution and energetics. We summarize our conclusions in \S 4. 

\begin{figure*}
\includegraphics[width=\textwidth]{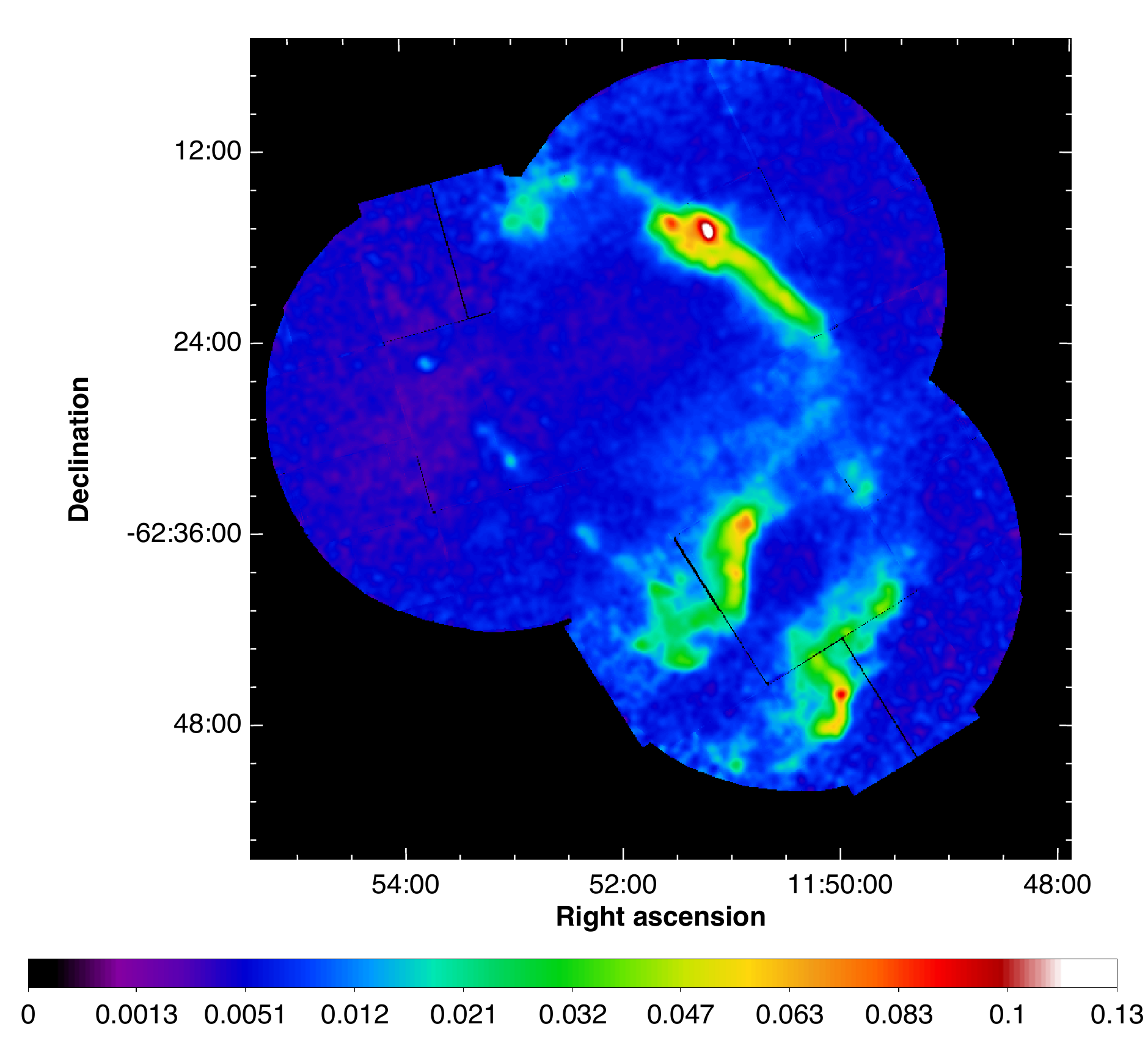}
\caption{EPIC MOS intensity map of G296.1--0.5 in the 0.3-3.0 keV energy band. A square root intensity scale is used and the units are photons/s/arcmin$^{2}$. The pixel binning is $4''$, and the image is smoothed with a Gaussian function of width $12''$. }
\label{fig:intensity}
\end{figure*}

\section{\emph{XMM-NEWTON} OBSERVATIONS AND ANALYSIS}

G296.1--0.5 was observed with \xmm\ during period A06. This set of observations covered the entire remnant with three pointings (see Table \ref{tab:obs}). Data from the European Photon Imaging Cameras (EPIC) PN, \mosi, and \mosii\ are analyzed using the \xmm\ Science Analysis System (SAS) version 9.0.0\footnote{The SAS package, and related documentation, is distributed by the \xmm\ Science Operations Center at http://xmm.esac.esa.int/sas/}, starting from the observational data files (ODFs). 

\input{tab1.tex}

For the MOS cameras, single to quadruple pattern events are selected, whereas for the data from the PN camera only singles and doubles are included in the analysis. Additionally, both for the imaging and spectral sections of the study, all flagged events are filtered from the event files. Proton flare contamination and periods of high background are removed by a procedure similar to that used in \citet{Arnaud01}. In the high energy band (10-12 keV for the MOS cameras, and 12-14 keV for the PN), the effective area of the cameras is small, and hence emission is dominated by the background. We create count rate histograms in these bands, and then remove periods where the number of counts per second exceed the standard limits (0.35 c/s and 4.0 c/s, for the MOS and PN cameras respectively). The effective exposure times after cleaning of the event files are shown in Table \ref{tab:obs}. The data from the SW pointing was highly degraded by flaring, and hence, in the imaging section of the analysis, the high energy count rate constraints are relaxed in order to allow for the morphology of the entire remnant to be studied. 

To complement the EPIC X-ray spectral study of \snr, we have also studied the remnant with the Reflection Grating Spectrometers (RGS 1/2), on board \xmm\ \citep{Herder2001}. The data from these instruments are analyzed with SAS version 9.0.0, starting from the ODFs, and the data selection and filtering are performed similarly to that of the MOS cameras. The RGS study is limited to the NW pointing, due to the degradation due to flares of the southwestern observation, and the lack of significant emission detected to the east of \snr.

\subsection{Imaging}

In order to analyze the morphology of G296.1--0.5, we create images of the entire SNR by combining the data from the three different pointings of the \mosi\ and \mosii\ cameras, using the SAS tasks \emph{merge} and \emph{evselect}. For each camera, events are selected in the energy bands 0.3-0.7, 0.7-1.0, and 1.0-3.0 keV. The images are created with a pixel bin size of 4$''$, and smoothed using a Gaussian with $\sigma=12''$(with the task \emph{asmooth}). Exposure maps for each energy band are obtained with task \emph{eexpmap}. The images are then combined, and divided by the exposure maps, into a single exposure-corrected intensity image, using the task \emph{emosaic}.

The \xmm\ EPIC MOS mosaic image of G296.1--0.5, in the 0.3-10.0 keV band, is shown in Figure \ref{fig:intensity}. The X-ray emission observed with \xmm\ describes a partial elliptical shell with three distinct lobes of emission, and corresponds well with the morphology observed with \rosat\ by \citet{Hwang1994}.  The brightest region is the northwestern shell, and there is a lack of significant emission in the east. Some faint diffuse emission is apparent northeast and southwest of the bright northwestern lobe.  

\begin{figure}[b]
\begin{center}
\includegraphics[width=\columnwidth]{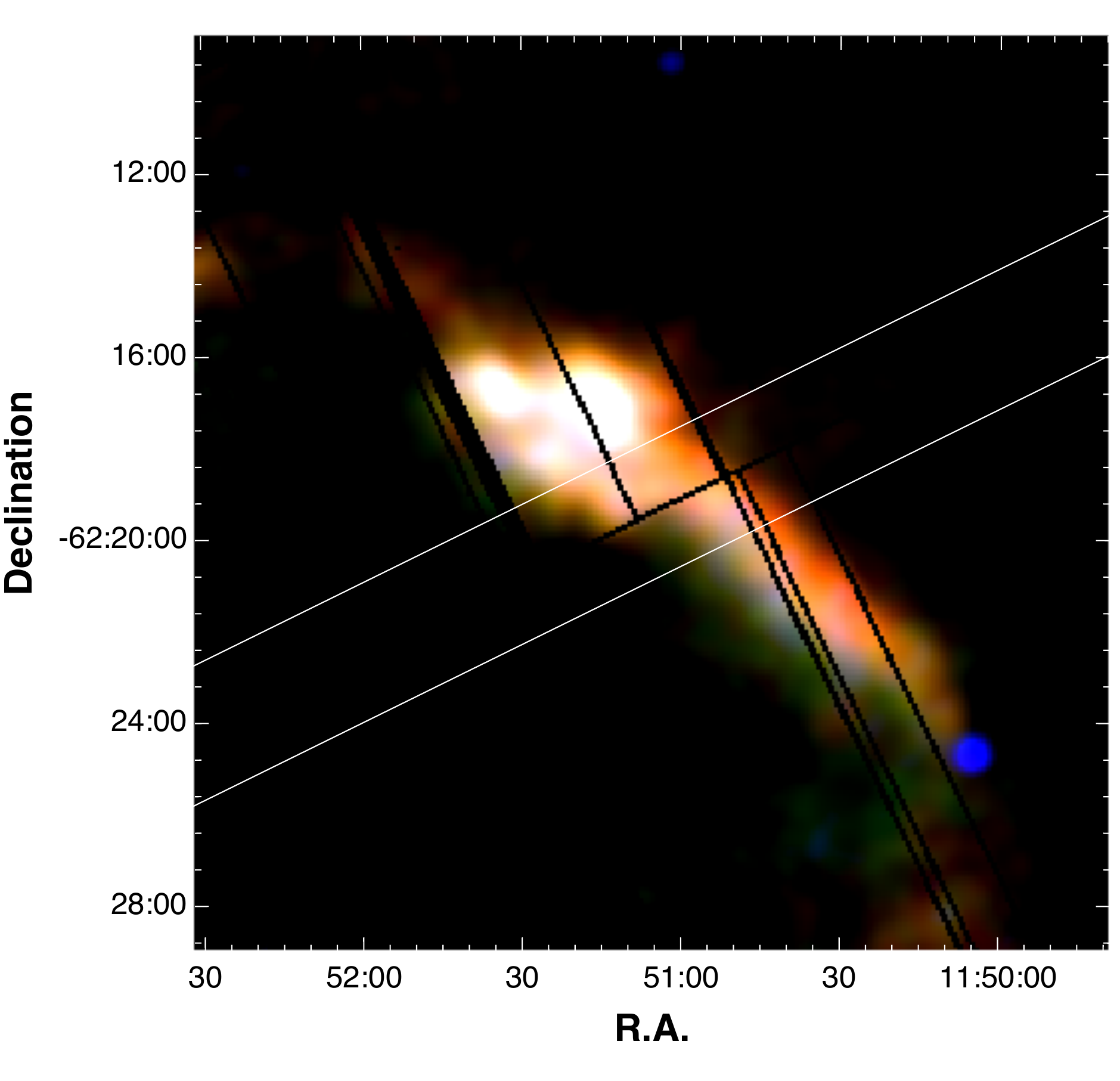}
\end{center}
\caption{Exposure corrected RGB image of the northwestern shell of SNR G296.1--0.5, created using EPIC PN data. Red corresponds to the 0.3-0.7 keV energy band, green to 0.7-1.0 keV, and blue to 1.0-3.0 keV. The image is $18'\times18'$, and smoothed with a Gaussian function of width 12$''$. The relative intensity levels of the color images have been adjusted to make the position of the hard compact source (SW of the shell) clear, and to highlight the broad spectral uniformity of the emission from the rim. The white lines indicate the area contributing to the RGS spectrum.}
\label{fig:rgb}
\end{figure}

With the aim of analyzing the structure of the northwestern limb of emission, and possible spectral variations within it, we also create an RGB image of this region, shown in Figure \ref{fig:rgb}, by combining images in the red (0.3-0.7 keV), green (0.7-1.0 keV), and blue (1.0-3.0 keV), from the EPIC PN camera data. This image reveals the morphology of this bright region as a clumpy rim, accompanied by a compact source (with a harder spectrum) that is discussed in detail in \S 2.3. In this RGB representation the relative intensity levels of the color images have been adjusted in order for the hard compact source to be clearly visible. 

\begin{figure*}
\begin{center}
\includegraphics[width=0.49\textwidth]{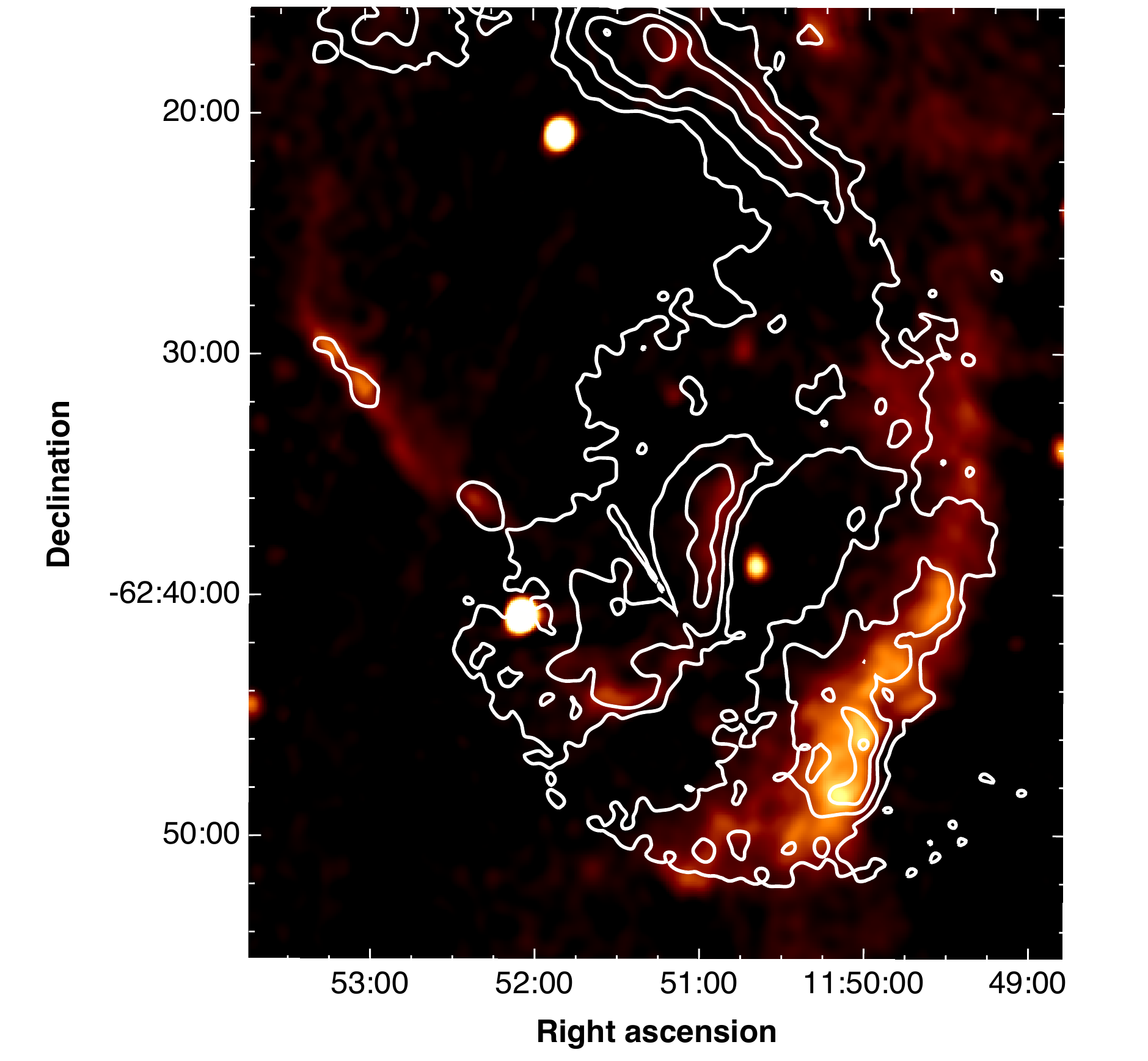}
\includegraphics[width=0.49\textwidth]{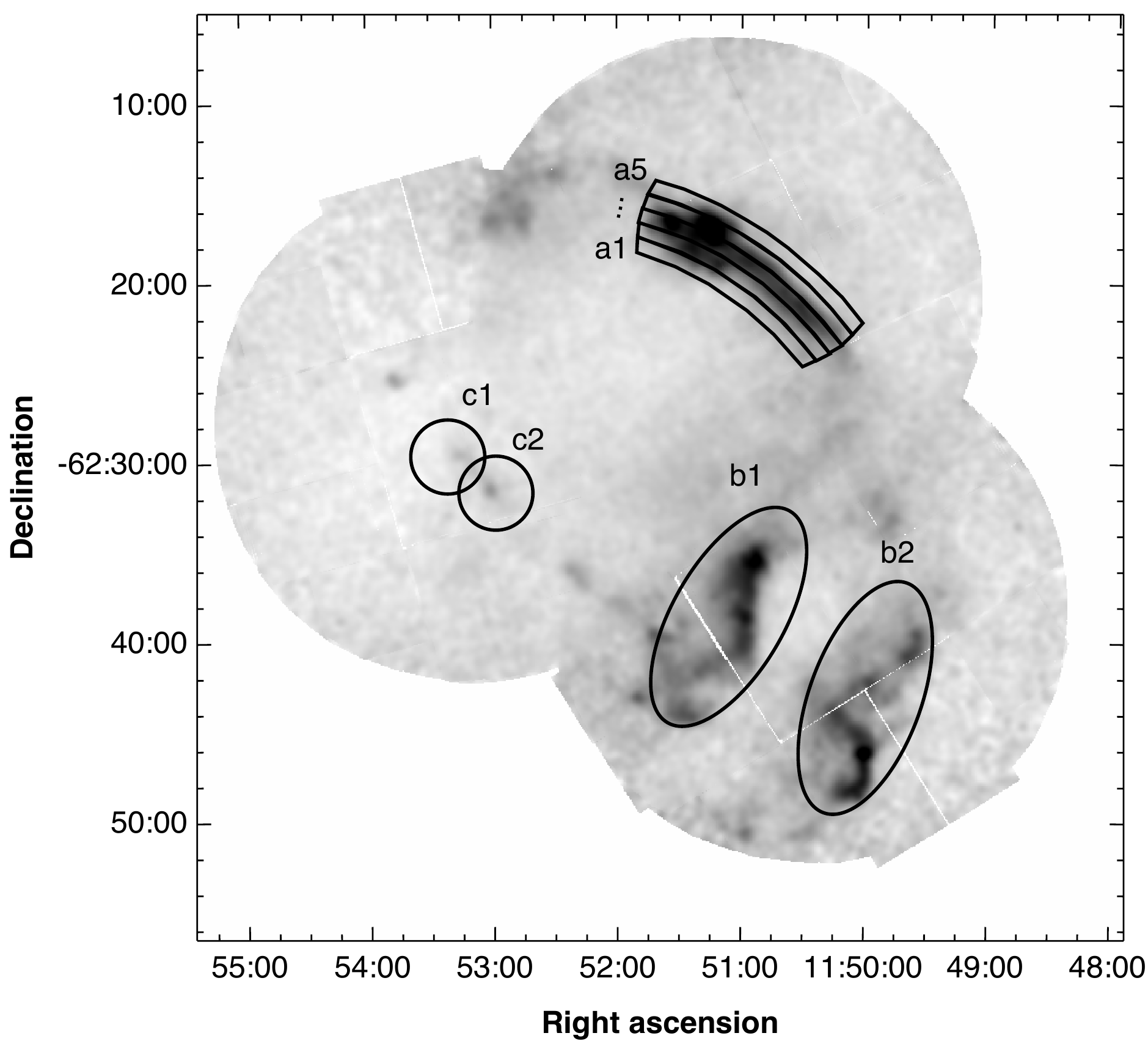}
\caption{\emph{Left}: MOST radio image at 1.4 GHz \citep{Whiteoak1996}. Overlaid in white are constant X-ray brightness contours from the EPIC MOS data that correspond to levels of 0.008-0.016-0.04-0.08 photons/s/arcmin$^{2}$.\emph{Right}: Mosaic image of the three pointings toward G296.1--0.5 in the 0.3-3.0 keV energy band, from the EPIC MOS cameras. The regions selected for spectral analysis are overlaid and labeled. }
\label{fig:reg}
\end{center}
\end{figure*}

The southwestern pointing was severely contaminated by soft proton flares (see Table \ref{tab:obs}), and hence a detailed morphological study of the double-limb structure in the SW is more difficult. As mentioned above, in order to study the morphology of this region of the SNR, the flare-filtering procedure is modified, allowing for a larger exposure time, but also higher background levels. Two distinct lobes of clumpy emission are clear, neither presenting clear shell structure, and faint diffuse emission dominates north of these (south and southwest of the northwestern shell). Additionally, we show the radio map of \snr\ obtained with the Molonglo Observatory Synthesis Telescope by \citet{Whiteoak1996} in Figure  \ref{fig:reg}(\emph{left}), overlaid with constant surface brightness contours from the EPIC MOS data. 


\newpage

\subsection{Spectroscopy}

In order to determine the spectral properties of the distinct areas of G296.1--0.5, spectra were extracted from the regions shown in Figure \ref{fig:reg}. The \xmm\ EPIC data of the northwestern part of the shell allow us to perform a detailed, spatially resolved, spectral study of this region. Broader aspects of the spectral properties of the two SW lobes, and the faint eastern region, are also studied. 

\xmm\ spectra are extracted from the cleaned event files using the SAS task \emph{evselect}. The detection efficiency of the \xmm\ mirrors is a function of off-axis angle, and is also energy dependent. Corresponding effective area and spectral response files for each of the extraction regions are created using tasks \emph{rmfgen} and \emph{arfgen}. The ancillary response files obtained with \emph{arfgen} allow for vignetting effects to be corrected. \emph{arfgen} is also used for calculating the BACKSCAL values for the extracted spectra, which represent the area of the source regions, corrected for CCD gaps and bad pixels.

The \xmm\ background is dominated by the Cosmic X-ray background (CXB) and the non X-ray background (NXB), which varies in time and results from a combination of the interaction of cosmic rays with the detector, background due to solar protons, and electronic noise \citep{Kuntz2008}. In order to account for the total \xmm\ background we use blank sky event files and Filter Wheel Closed (FWC) data provided by the \xmm\ EPIC Background Working Group\footnote{Filter Wheel Closed and blank sky data from the \xmm\ EPIC Background Working Group are available at http://xmm2.esac.esa.int/external/xmm\_sw\_cal/background/}. The blank sky data, obtained using observations of several sky positions, is a combination of the CXB and the NXB \citep{Carter2007}. Because the NXB changes in time, the blank sky event files are combined with the FWC data in order to match the instrumental background levels between our observations and the background data sets, as explained in \citet{Temim2009}. In this analysis the background subtraction is performed through a two step process which uses both corrected blank sky products, and data from parts of our observations outside the source regions, similar to the method detailed in \citet{Arnaud2002}. 

The spectra were analyzed with the XSPEC software package (version 12.5.1), using an energy range of 0.3-2.0 keV, grouped with a minimum of 25 counts bin$^{-1}$, and using the $\chi^2$ statistic. The emission above $\sim 2$ keV is dominated by the background. In all regions, the model that best describes the spectral characteristics is that for the absorbed emission from a non-equilibrium ionization collisional plasma, with variable abundances, VNEI \citep{Borko2001}. The parameters of the best-fit models for all extracted spectra are shown in Table \ref{tab:mosno2}, where the uncertainties quoted are the 90\% confidence limits (1.6$\sigma$). 

\input{tab2.tex}

\begin{figure}[b]
\includegraphics[width=\columnwidth]{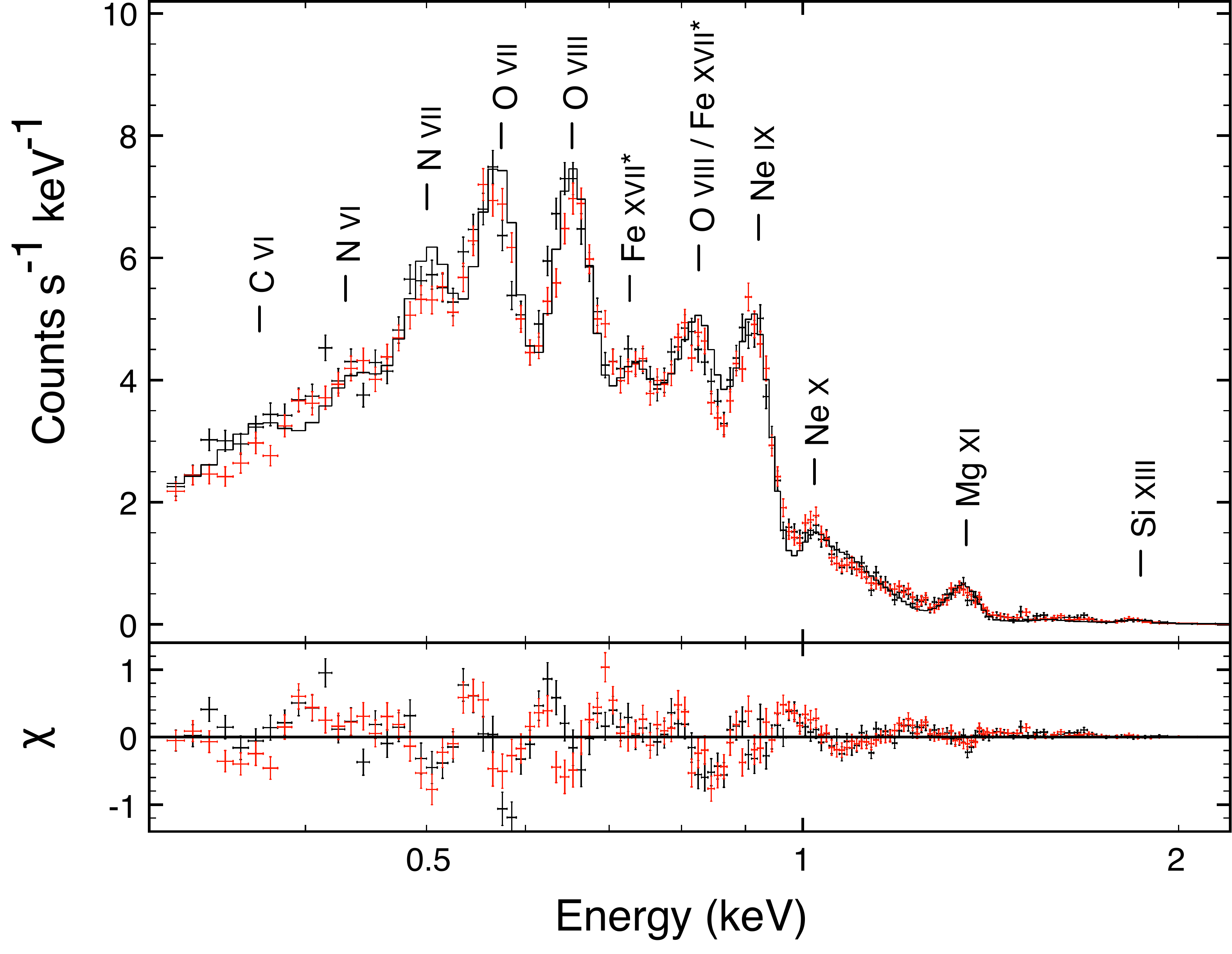}
\caption{\xmm\ EPIC MOS 1/2 background subtracted spectra of the northwestern shell of \snr\,  with VNEI model fits. The positions of the C {\scriptsize VI}, N {\scriptsize VI}, N {\scriptsize VII}, O {\scriptsize VII}, O {\scriptsize VIII}, Ne {\scriptsize IX}, Ne {\scriptsize X}, Mg {\scriptsize XI}, and Si {\scriptsize XIII} emission lines are indicated. The lines marked with an asterisk are those for which definitive identification is beyond the scope of this work. Above 2 keV there is no significant signal in the background subtracted spectrum.}
\label{fig:a_spec}
\end{figure}

The integrated \mosi\ and \mosii\ spectra for the entire northwestern shell (with VNEI model fits) are shown in Figure \ref{fig:a_spec}. The best-fit absorbing column density is low,  \nH\ =$2.00^{+0.73}_{-0.83} \times 10^{20} \text{cm}^{-2}$, and the plasma temperature is $0.63\pm 0.04$ keV. The emission line features corresponding to N {\footnotesize VII}, O {\footnotesize VII}, O {\footnotesize VIII}, and Ne {\footnotesize IX} are clear, as well as the Ne {\footnotesize X}  Ly$\alpha$ line,  and the helium-like Mg {\footnotesize XI} unresolved triplet. Emission lines possibly indicating Fe L-shell transitions are also apparent, but the contribution from K$\beta$ lines from oxygen is difficult to estimate precisely, and a definitive identification of spectral features in this energy region falls beyond the scope of this work. Abundances of C, Si, S, Ar, Ca, Fe  and Ni are fixed at solar values, taken from \citet{Anders1989}. The Ne, N, and O abundances are not consistent with solar, and the fit improves significantly when these values are allowed to vary.  Regions a1 through a5 are subsections of the northwestern lobe, and the best-fit model parameters for these are presented in Table \ref{tab:mosno2}. All individual model fits are consistent with the column density obtained for the full northwestern shell, and were hence fixed at that value. Enhanced abundances of N (a factor of 2 greater than solar values, approximately),  and underabundances of O, are favored in the fits to the spectra of all regions, as well as ionization timescales in the range 2-4 $\times 10^{10}$ s cm$^{-3}$. The abundance of Mg in most regions appears to be consistent with the solar value within the uncertainties obtained, and it represents another common characteristic of the extracted spectra across all regions studied.

The spectra extracted for the southwestern limbs of emission, regions b1 and b2, show similar characteristics to those from the northwestern shell. The temperatures are $\sim 0.6$ keV, and the relative abundances and ionization timescales are also consistent. The spectra from the eastern part of the remnant, regions c1 and c2, do not display significant divergence from the overall fit, but the faintness of the emission in this area makes statistically significant conclusions difficult to establish.

\begin{figure}[t]
\includegraphics[width=\columnwidth]{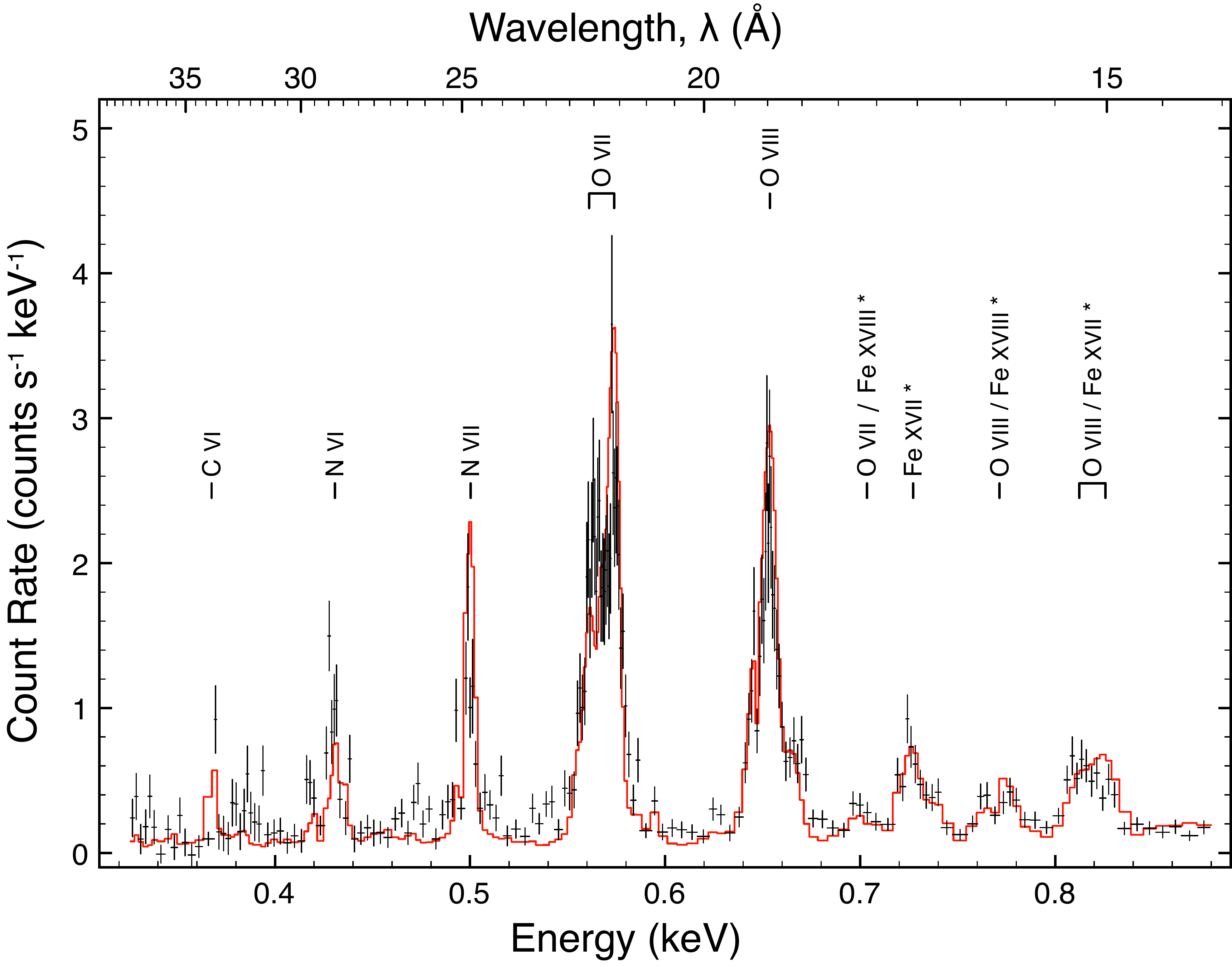}
\caption{\xmm\ RGS background subtracted spectrum of a small region in the northwestern shell of \snr.  The VNEI emission model obtained by fitting the EPIC spectra is shown as a histogram. The positions of the C {\scriptsize VI}, N {\scriptsize VI}, N {\scriptsize VII}, O {\scriptsize VII}, and O {\scriptsize VIII} emission lines are indicated. The lines marked with an asterisk are those for which definitive identification is beyond the scope of this work.}
\label{fig:a_rgs}
\end{figure}

A high resolution spectrum of a section of the NW limb, obtained with the RGS 1 instrument, is shown in Figure \ref{fig:a_rgs}. The extraction region selected is a 2.8$'$ wide strip across the northwestern shell, centered at $\alpha_{\text{2000}},\delta_{\text{2000}}=11^{\text{h}}50^{\text{m}}56.748^{\text{s}}, -62^{\circ}18'52.25''$, and it is shown on Figure \ref{fig:rgb}. The background is calculated using the SAS task \emph{rgsbkgmodel}, which uses linear combinations of background templates from empty fields and information from the observation of interest, to derive the applicable model background \citep{Gonzalez2004}. The emission model shown as a histogram in Figure \ref{fig:a_rgs} is the same absorbed VNEI model used to fit the EPIC spectra of the entire NW shell, renormalized to account for the smaller number of counts in the RGS spectrum. Since the region studied is extended, with an approximate spatial width of 3$'$  (FWHM), the model has been convolved with the spatial image of the area in order to fit appropriately the spectrum, using the XSPEC (version 11.3) convolution model \emph{rgsxsrc}\footnote{This code was developed by Andy Rasmussen of the Columbia University XMM/Newton RGS instrument team, and it convolves the spectral model of interest with an angular structure function taken from an appropriate image of the extended source. To fit the data, the resulting convolved spectral model is used with the standard RGS point source spectral response function.}, which results in an effective spectral broadening $\Delta E/E\sim0.02$. While the extension of the region and low statistics make it difficult to use the RGS data to obtain stronger constrains on the abundances and ionization state of the emitting plasma in the NW shell, the RGS 1 spectrum shown in Figure \ref{fig:a_rgs} clearly show emission lines of C {\footnotesize VI}, N {\footnotesize VII}, O {\footnotesize VII}, O {\footnotesize VIII}, Fe {\footnotesize XVII}, and appears suitably described by the emission model derived from the EPIC observations. 

This spectral analysis differs greatly from that of \citet{Hwang1994} for \rosat\ PSPC observations of \snr, where the spectrum is fit to a model for a plasma in ionization equilibrium, resulting in lower temperature, higher interstellar absorption, and very low elemental abundances relative to solar, as noted in \S~1. In order to understand this discrepancy, we first attempted to fit the MOS spectrum to the best-fit model obtained by \citet{Hwang1994}. We find that the model is rejected at high significance, yielding a reduced  $\chi^2$ statistic of 7.8. We then re-extracted the \rosat\ PSPC spectrum and investigated fits to our model results. The absorbed NEI plasma model provides a very successful fit to the PSPC spectrum, and requires neither a second temperature component, nor uncommon abundances relative to the solar values. We find that the \rosat\ data prefer a somewhat higher column density (\nH\ $\approx 8\times10^{20}\text{cm}^{-2}$) and lower temperature ($kT\approx0.35$ keV), after which the model yields a reduced $\chi^2$ statistic of 0.95. This appears to be a result of the complexity of spectrum of the plasma, and the low spectral resolution of the \rosat\ data set. We note that the low temperature and higher column density associated with the fits obtained by \citet{Hwang1994} result in the bulk of the inferred emission being absorbed. To match the spectrum at higher energies, a very large normalization is required, resulting in a very large inferred mass, contrary to what our fits indicate (see below, Section 3.2).

\begin{figure*}
\begin{center}
\includegraphics[width=0.49\textwidth]{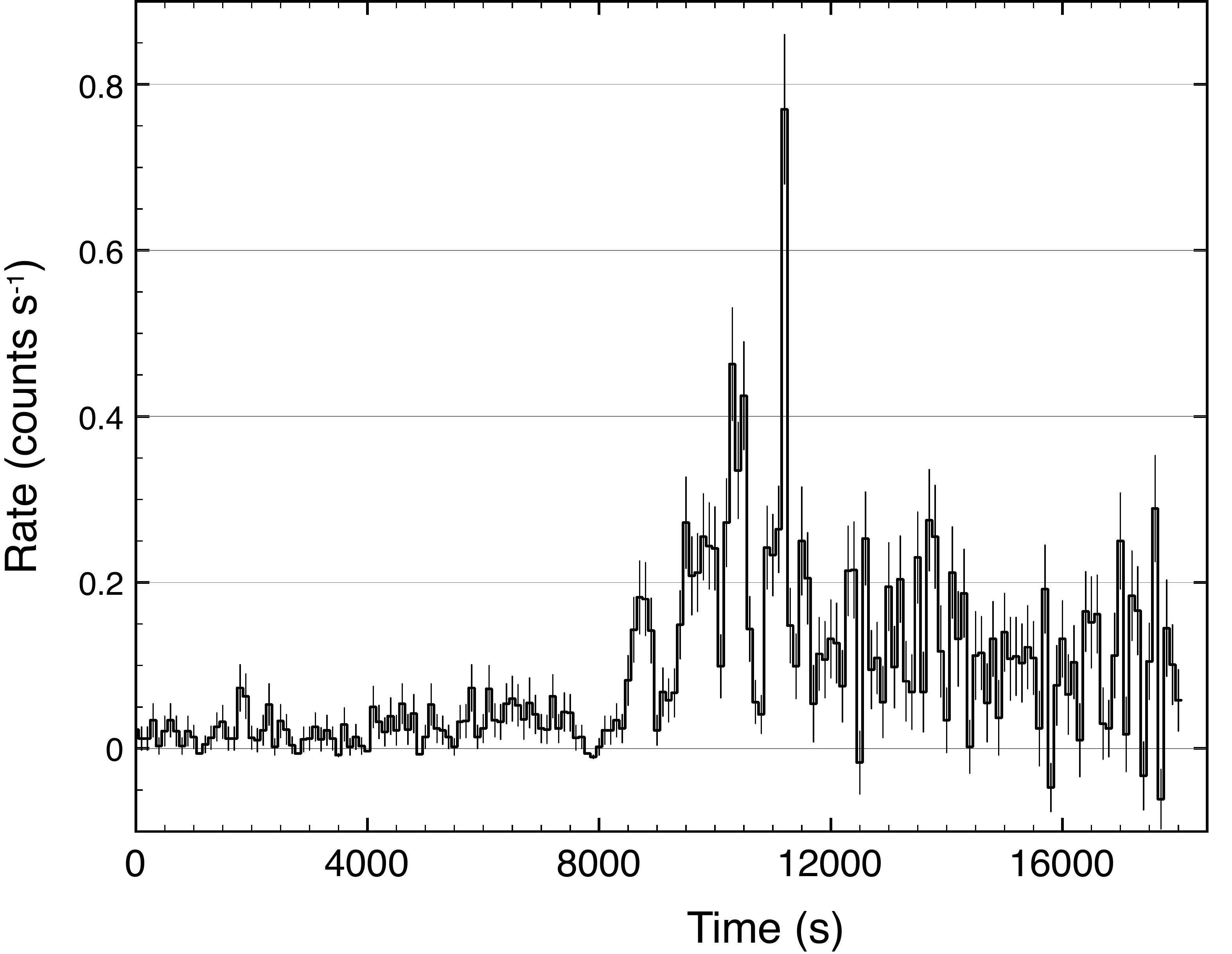}
\includegraphics[width=0.49\textwidth]{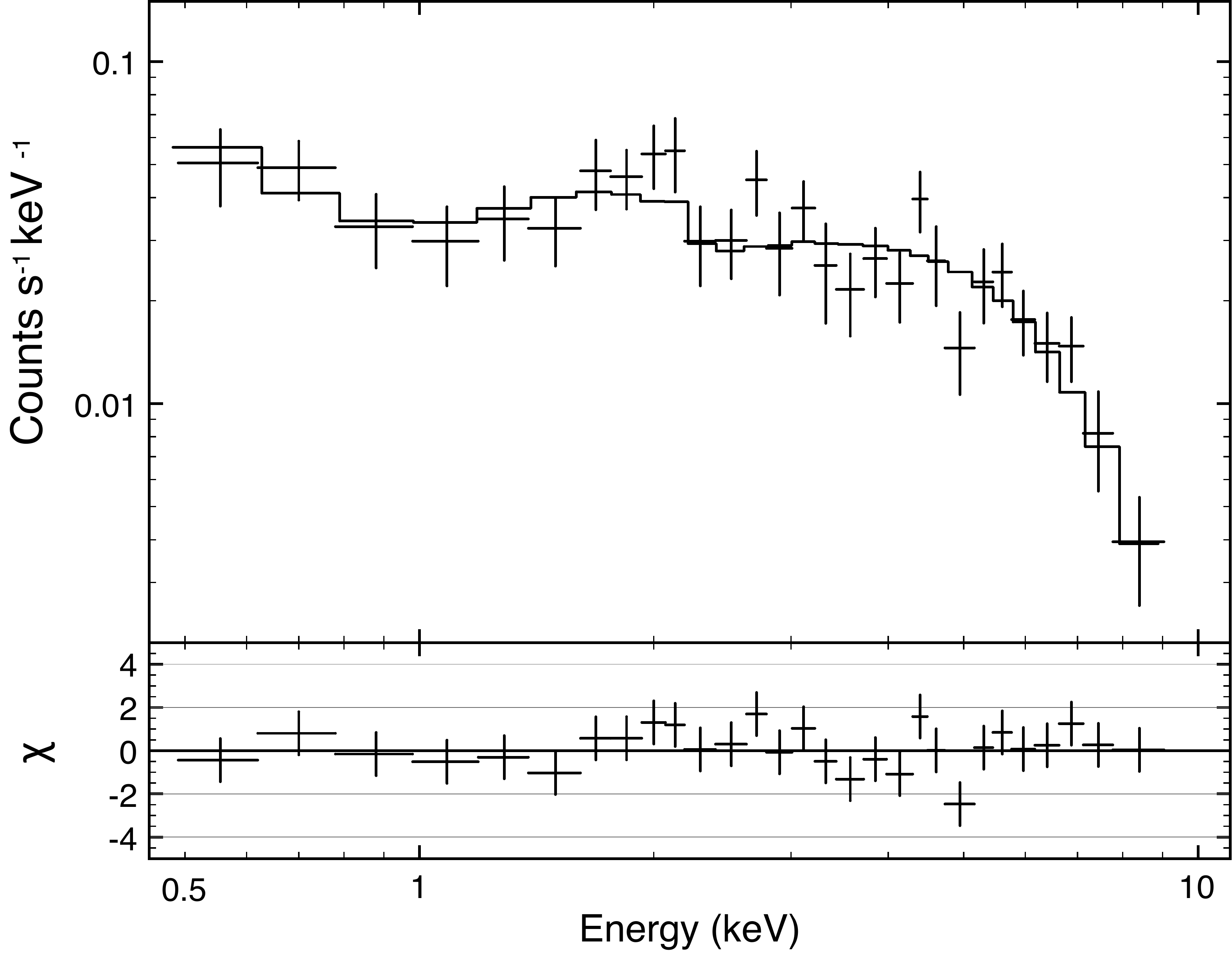}
\caption{\emph{Left}: Background subtracted light curve of \ps\ during the full exposure interval. The background level rises significantly at t$>12,000\,\text{s}$, and hence the uncertainties on the count rate values increase as well, making it difficult to establish if the source fades below detectability. \emph{Right}: Spectrum during the burst period (see \S 3.3), from the EPIC PN data in the 0.3-10 keV energy band. The histogram in the spectral distribution corresponds to the best-fit spectral model, that of a blackbody combined with a power law (see model parameters in Table \ref{tab:ps}).}
\label{fig:ps_prop}
\end{center}
\end{figure*}

\subsection{A Compact Source Near the NW shell}

Of additional interest from our study of G296.1--0.5 is the discovery of a bursting X-ray source,  at $\alpha_{\text{2000}},\delta_{\text{2000}}=11^{\text{h}}50^{\text{m}}05^{\text{s}}, -62^{\circ}24'43''$, near the northwestern limb of the remnant. The absolute astrometry of \xmm\ is 2$''$, and the statistical uncertainty is dependent on the intensity of the source, hence limiting the 95\% confidence contours to $\sim5''$ \citep{Kirsch2004,Kong2007,Guainazzi2010}. This source is listed in the Second XMM-Newton Serendipitous Source Catalog, Third Data Release (2XMMi-DR3) as  2XMMi~J115004.8--622442 \citep{Watson2009}\footnote{The 2XMMi-DR3 catalog, released by the European Space Agency on 2010 April 28, can be obtained from the XMM-Newton Survey Science Center (http://xmmssc-www.star.le.ac.uk).}. In Figure \ref{fig:rgb}, the source 2XMMi~J115004.8--622442, henceforth \ps, is clearly visible as a compact source of hard emission located just SW of the northwestern lobe. 

The left panel in Figure \ref{fig:ps_prop} shows the background subtracted EPIC PN light curve of \ps\ in the 0.3-10 keV band, for the full exposure interval. The source is not detectable during the initial 8~ks of the observation, and then undergoes a burst lasting approximately 3.8~ks, with a net count rate 0.2 counts s$^{-1}$. The burst appears to present some substructure with scales of 500 s to 1000 s. The outburst is seen in both the \mosii\ and PN cameras. (The source falls on the non-functioning chip on \mosi). Just as the burst begins to decline, the soft proton flare rate in the detectors begins to climb precipitously and hence the uncertainties on the background subtracted count rates increase. While we can clearly establish that the source flux declines significantly after the burst, it is not clear whether or not it fades all the way to the level of undetectability, as in the first part of the observation. We searched for a periodic signal for periods between 147 ms and 200 s (the lower limit is due to the time resolution of the PN camera, and the upper limit is set considering the usual periods of pulsars and magnetars in X-rays), and detected  no evidence for pulsations. At the 99\% confidence level, the upper limit on the pulsed fraction for a sinusoidal modulation is 45\%.


The time-averaged spectrum of the source during the burst period (defined as $8,029\,\text{s}<t_{\text{burst}}<11,855\,\text{s}$) is shown in the right panel of Figure \ref{fig:ps_prop}. The spectrum bins are grouped in order to have a minimum 25 counts per bin, and hence the lowest energy bin is centered on 0.55 keV. Not unlike \snr, the spectrum of \ps\ is best-fit using models with low column density, \nH\ =$0.07^{+0.47}_{-0.05} \times 10^{22} \text{cm}^{-2}$. The statistical uncertainties are large, yet this estimate suggests a distance that is consistent with that of the remnant. Single-component models (thermal or nonthermal) do not provide a good fit to the spectrum, but a variety of two-component models can describe the data. The best-fit model is that of a blackbody combined with a power law, with model parameters as shown in Table \ref{tab:ps}. It is not possible to study the time-evolution of the spectrum due to the low number of counts detected during the outburst.

\input{tab3.tex}

\section{DISCUSSION}

\subsection{Distance Estimate}

The optical study of \snr\ by \citet{Longmore1977} estimates the distance to the SNR using reddening measurements, as well as by fitting the radial velocity observed, with respect to the local standard of rest (LSR), to Galactic rotation. Using H$\alpha$ and H$\beta$ observations, and reddening measurements in the direction of Crux \citep{Miller1972}, these authors estimate the distance to \snr\ to be 3$\pm1$ kpc. The radial velocity of the optical filaments observed is -35 km s$^{-1}$, consistent with H$\alpha$ observations by \citet{Russeil2001}, which places the SNR at the tangent point, 3.8 kpc distant (using a distance of 8.5 kpc to the Galactic center). The velocity error of 3 km s$^{-1}$, combined with the systematic error associated with the uncertainties in the rotation curve and with non-circular motions (7 km s$^{-1}$), corresponds to a distance uncertainty of $\pm1.9$ kpc. 

Using CO observations, \citet{Brand1987} also detected molecular material with consistent radial velocities to those of the optical filaments observed by \citet{Longmore1977} ($\sim-38$ km s$^{-1}$) at $\alpha_{\text{2000}},\delta_{\text{2000}}=11^{\text{h}}50^{\text{m}}27^{\text{s}}, -62^{\circ}38'55''$. The position of this molecular material is coincident with the faintest area of both radio and X-ray emission between the S and SW lobes of emission. 

The measurements of the X-ray absorption from the spectral analysis of all regions of the SNR in the \xmm\ observation indicate low column densities, which suggests that the distance to the object is small. This result is consistent with the lower limit of the distance estimate ($\sim$2 kpc) obtained from the optical observations, and we scale subsequent quantities here with $d_2 = d/(2 {\rm\ kpc}$). The total unabsorbed flux (in the energy band 0.3-10.0 keV) for the northwestern lobe is 2.88$\times 10^{-11}$ erg cm$^{-2}$ s$^{-1}$, and implies a luminosity of $1.4\times 10^{34}d_2^{2}$ erg s$^{-1}$. 

Assuming such distance, $d=2$ kpc, the column density estimated from the spectral analysis of the X-ray data implies that the average hydrogen density along the line of sight is $<n_{\text{H}}>\sim$0.03 cm$^{-3}$, which is low relative to the average Galactic density. We note that \citet{McClure2001} detected a giant H {\footnotesize I} shell, GSH 304--00--12, at a distance $d\sim1.2$ kpc, which is located in front of \snr. This suggests that the low foreground column density derived from the X-ray observations is partly the result of observing the SNR through this giant shell, which has swept some of the gas away from the line-of-sight of the remnant.

\subsection{Possible Origins and Evolutionary Characteristics}

The chemical properties of the emitting material appear to be consistent across the different X-ray bright regions, like most of the other spectral characteristics. We find the nitrogen enhancement (by
a factor of 1.5 to 2 in comparison to solar values) and the oxygen deficit (by a factor of approximately 1.5-2) are of particular interest, and appear to persist from region to region. Nitrogen enrichment and oxygen depletion are typical of products from the CNO cycle, and are often observed in winds from red supergiant (RSG) stars, and from the subclass WN of Wolf-Rayet stars \citep{Esteban1992,Garcia1996,Crowther2007}. We suggest that the spectrum indicates that \snr\ is the result of a core-collapse supernova, and that the SNR is currently expanding into the late-phase wind of its massive ($> 25 M_\odot$) progenitor. The nitrogen enhancement observed in the CSM of these massive stars is usually greater than measured in these observations of \snr, with enrichment factors larger than 3 \citep{Garcia1996}, and hence some dilution of the CNO process products appears to be required.

\citet{Garcia1996} investigate the characteristics of the circumstellar medium (CSM) around massive stars that are subject to severe mass loss. The properties of the stellar winds from massive stars depend on the phase of stellar evolution. During the RSG phase, the mass loss rate is high (almost 10$^{-4}$ $M_{\odot}\text{yr}^{-1}$), and the terminal wind velocity, $v_w$, is less than 100 $\text{km}\,\text{s}^{-1}$. The stellar wind is much faster, of the order of several 1000 $\text{km}\,\text{s}^{-1}$, in the Wolf-Rayet (WR) phase, and the mass loss rate is $\dot{M} \approx 2-3 \times 10^{-5}\,M_{\odot}\text{yr}^{-1}$. The fast WR wind sweeps up the dense wind from the RSG phase, and at the time of the supernova, a low density wind bubble is expected to have formed with a shell of clumped wind material at a radial distance $\gtrsim10$ pc, with density $\rho\sim10^{-25}$ g cm$^{-3}$. Stellar evolution, and observations of nebulae formed around WR stars, suggest that the composition of the shell of material swept-up by the wind should be overabundant in N, and O deficient \citep{Esteban1992,Chevalier2005}.

The prevailing theoretical model suggests that single stars with mass $\sim 25-35$ $M_{\odot}$ end their lives as red supergiant stars with mass-loss rates $\dot{M}\gtrsim 3\times 10^{-5}$ $M_{\odot}\text{yr}^{-1}$ for $v_w=15$ $\text{km}\,\text{s}^{-1}$, and result in supernovae (SNe) type IIL/b \citep[and references therein]{Chevalier2005}. Stars of mass $\gtrsim 35$ $M_{\odot}\text{yr}^{-1}$ are expected to explode in supernova explosions type Ib/c, after having entered the WR phase. The circumstellar environment of WR stars can be extremely complex, and specific hydrodynamic models are needed to follow the mass-loss evolution before the explosion, and the subsequent supernova interaction with the CSM \citep[e.g.][]{Dwarkadas2007}. However, the case of an SN  IIL/b interacting with the wind of the progenitor RSG star can be studied in an analytical fashion, and the following discussion considers the observations of \snr\ in such a scenario.

In order to derive the evolutionary properties of \snr\ we use the observed characteristics of the remnant and the model treatment for an SNR expanding into an RSG wind prescribed by \citet{Chevalier2005}. The CSM is presumed to have the profile of a freely expanding wind, with density $\rho_{\text{CS}}=\dot{M}/4\pi r^2v_w\equiv Dr^{-2}$. For characteristic properties of an RSG wind, $\dot{M}= 3\times 10^{-5}$ $M_{\odot}\text{yr}^{-1}$ and $v_w=15$ $\text{km}\,\text{s}^{-1}$, the value of the coefficient of the density profile is $D_{\text{ch}}=10^{14}$ g cm$^{-1}$. Hence, \citet{Chevalier2005} defines $D_{\ast}=D/D_{\text{ch}}$, and establishes that the circumstellar mass swept up by the SNR shock, at a radius $R$, is given by 
\begin{equation}
M_{\text{sw}}=9.8D_{\ast}\left(\frac{R}{5\text{ pc}}\right)\, M_{\odot}\text{.} 
\end{equation}

\noindent \citet{Chevalier2005} finds that, for a given explosion energy $E_{\text{SN}}$, the forward shock of an SNR in a wind, in the adiabatic phase of its evolution expands as

\begin{equation}
R = \left(\frac{3E_{\text{SN}}\, t^2}{2\pi D}\right)^{1/3}\text{.}
\end{equation}
%
%
%
%

The estimated angular radius of \snr\ is 20$'$, assuming the NW lobe and outer southern limb form the shell of the remnant, and implies a remnant radius $R_{\text{SNR}}\approx12\,d_2$ pc. For a strong shock propagating in a wind with density profile $\propto r^{-2}$, the shocked CSM is swept up into a shell approximately $R/5$ in thickness \citep{Chevalier1982}. Hence, from the fitted volume emission measure of the X-ray spectrum of the NW shell, $EM\approx1.2\times10^{56}d_2^{2}$ cm$^{-3}$, the total swept up circumstellar mass is $M_{\text{sw}}\approx$19$\,d_2^{5/2}\, M_{\odot}$. 
This value assumes that the observed X-ray emission is entirely associated with the shocked CSM, that the SNR has a spherical shell geometry with radius 20$'$ and thickness 4$'$, and that we observe only a fraction 4.8\% of the SNR in the NW shell.
If the SNR is indeed expanding into an RSG wind, then the swept up mass and radius can be used in Equation 1 to obtain an estimate of the coefficient of the density profile, $D=8\times 10^{13}d_2^{3/2}\,$ g cm$^{-1}$. Using Equation 2, we can infer the age of the SNR to be $t\approx 2800\,E_{51}^{-1/2}\,d_2^{9/4}$ years, where $E_{51}=E_{\text{SN}}/(10^{51}\,\text{erg})$. The estimates of the ionization timescale, $n_{\text{e}}t$, obtained from the spectral analysis of all regions of the remnant (as shown in Table \ref{tab:mosno2}) appear to be approximately $2\times 10^{10}$ s cm$^{-3}$, which, combined with the age derived, suggests that the electron density is 0.22$E_{51}^{1/2} d_2^{-9/4}$ cm$^{-3}$, which is in excellent agreement with the value of $0.17d_2^{-1/2}$ cm$^{-3}$ inferred from the emission measure. For an explosion energy of $E_{51}=1$, these independent estimates of the electron density yield the same value of 0.15 cm$^{-3}$ if the SNR is at a distance of 2.4 kpc. 
Using Equation 2, for the expansion of an SNR in a wind density profile, we can derive an expression for the speed of the forward shock at any given time, $V_{\text{S}}=2R/(3t)$. The shock velocity that results from the evolutionary description of \snr\ is $V_{\text{S}}=2700 \,E_{51}^{1/2}\,d_2^{-5/4}$ km s$^{-1}$. The derived values under the assumed RSG evolution should be contrasted with the values of $t\approx 2100\,E_{51}^{-1/2}\,d_2^{9/4}$ years, and $M_{\text{sw}}\approx15\,d_2^{5/2}\,M_{\odot}$ that are obtained under the assumption of a Sedov solution \citep{Sedov1959,Taylor1950}.

\subsection{Clumped Shocked Material and an SN Type Ib/c Origin}

As described in Section 4.2, SNe Types Ib and Ic are expected to originate from the core-collapse of Wolf-Rayet (WR) stars \citep{Chevalier2005}. In this scenario, the supernova shock will expand quickly through the low-density bubble, and then interact with the denser shell of wind material. Observations in the radio and X-ray bands of SNR Kes~75 suggest that the high shock velocity and young shock age of the reverse-shocked ejecta are consistent with this remnant being the result of the collapse of a WR star \citep{Chevalier2005,Morton2007}. Using radio observations of SNR G320.4--1.2, and its associated pulsar PSR B1509--59, \citet{Gaensler1999} argue that the remnant is the result of a Type Ib/c supernova explosion. \citet{Seward1983} observed optical knots in SNR G320.4--1.2 at a radius $\gtrsim10$ pc, with strong [\ion{N}{2}] lines, and \citet{Chevalier2005} suggests the characteristics of these knots support the scenario where G320.4--1.2 is the result of an SN Ib/c event. The clumpy nature of the N-enriched CSM indicated by the X-ray emission from \snr, may indicate that this remnant also originated from the collapse of a WR star.


Measurements of the strength of electron scattering wings in WR emission lines, as well as the shape of the IR and radio continua, provide strong evidence for clumping in the wind \citep[and references therein]{Willis1999}, with filling factors as large as $10 - 20\%$ \citep{Lepine2000,Morris2000}.  Such clumping may result from instabilities in the radiatively-driven winds. Although the size-scale and density contrast of the clumps observed within the bright X-ray emission regions is somewhat uncertain, models by \citet{Garcia1996} result in structures in the wind with solid angles $\Delta \Omega \sim 0.05$, which remain constant with expansion radius. If maintained out to radii as large as \snr, these structures would have sizes similar to the clumpy structure that we appear to observe in the X-ray shell. The large-scale clumping seen in Figure \ref{fig:rgb} is clearly real, and approximately half a dozen of the smaller structures represent $3 - 5 \sigma$ brightness enhancements relative to the average emission in the shell.  


\citet{Crowther2007} notes that WR stars are located within, or nearby, star-forming regions in the Galactic disk. The environment of \snr\ is consistent with such a situation since \citet{Brand1987} detected a nearby compact  H {\footnotesize II} region, G295.69--0.34, using CO observations. This nebula is located $\sim11'$ west of the edge of the NW shell, at $\alpha_{\text{2000}},\delta_{\text{2000}}=11^{\text{h}}50^{\text{m}}18^{\text{s}}, -62^{\circ}28'10''$. The radial velocity of this region is $V_{\text{LSR}}\sim-18\text{km}\,\text{s}^{-1}$, which corresponds to a distance of 2-5 kpc, by fitting to Galactic rotation curves. The properties of this nebula lead to its identification as a star forming region, and it is hence catalogued as such by \citet{Avedisova2002}. Additionally, \citet{Shara1999} detected WR star 45b at $\alpha_{\text{2000}},\delta_{\text{2000}}=11^{\text{h}}48^{\text{m}}46^{\text{s}}, -62^{\circ}23'03''$, within 10$'$ west of the NW shell. These authors constrain the distance to this WR star to be $d\leq8.7$ kpc, which does not rule out its association with the star-forming region  mentioned above. Thus, while a detailed investigation of the expected X-ray emission under such a scenario requires hydrodynamical simulations beyond the scope of this paper, the observed properties of \snr\ may be consistent with a WR progenitor.

\subsection{On the nature of the compact source}

\begin{figure}
\begin{center}
\includegraphics[width=\columnwidth]{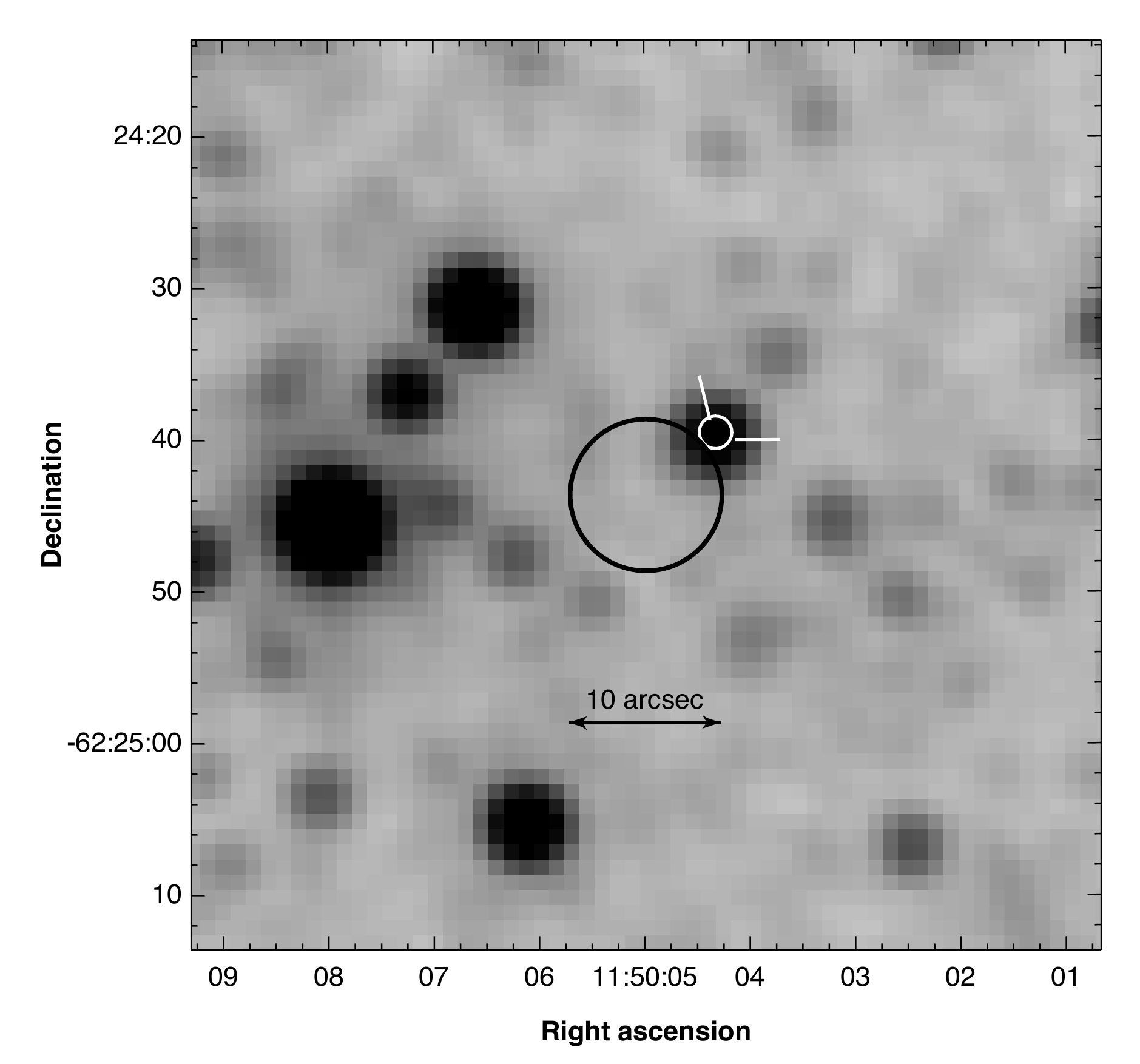}
\end{center}
\caption{2MASS  image in the \emph{J}~band toward \ps. The field shown is $1'\times1'$, and the black circle indicates the position of the \xmm\ source with a 5$''$ radius, which represents the absolute astrometry uncertainty of the instrument. The white circle is centered on the position of the source as detected by \emph{Chandra}, and the radius of the region is 1$''$ (approximately the uncertainty on the position). The two tick marks indicate the position of 2MASS~J11500433--6224398.}
\label{fig:ps_2mass}
\end{figure}

There are a wide range of X-ray sources that show evidence of variability similar to that observed from \ps, particularly the burst's apparent duration ($\sim 4000 s$), like stars undergoing flaring periods, X-ray bursts from binary systems and magnetars. The absolute astrometry uncertainty of the \xmm\ EPIC cameras does not allow to confirm the possible association between this bursting source and nearby source 2MASS~J11500433--6224398, observed in the 2MASS and DSS images (located about 5 arcseconds from the X-ray centroid -- see Figure \ref{fig:ps_2mass}). However, a recent observation with the \emph{Chandra} Advanced CCD Imaging Spectrometer (ACIS-I) detector on 2010 October 12 for $\sim30000$ ks (ObsID 12560), with sub-arcsecond spatial resolution, places \ps\ at effectively the same position ($\alpha_{\text{2000}},\delta_{\text{2000}}=11^{\text{h}}50^{\text{m}}04.32^{\text{s}}, -62^{\circ}24'39.5''$) as 2MASS~J11500433--6224398, suggesting this is in fact the counterpart of the bursting source. This IR source has $J/H/K$ magnitudes $11.93\pm0.03$, $11.44\pm0.02$, and $11.30\pm0.02$ respectively \citep{Skrutskie2006}. The X-ray variability, and spectral characteristics, together with the possible association with the SNR, suggest \ps\ could be a member of the magnetar family. Given the position of this source, and suggestions that magnetars arise from unusually massive progenitors \citep{Gaensler2005}, it is tempting to suggest an association between \ps\ and SNR \snr. Magnetars 1E~2259+586 and 1E~1841-045 have been associated with core collapse SNRs CTB~109 and Kes~73, respectively \citep[and references therein]{Woods2006}. However, the IR counterpart appears to rule out the magnetar interpretation.

Based on the variability, it is possible that this source is a low mass X-ray binary (LMXB) undergoing a Type I X-ray burst, as accreting material falling into a neutron star fuels a thermonuclear flash. The spectrum of such events is well modeled by blackbody radiation with temperature $T\approx1-2$ keV, which is consistent with the best-fit model to the spectrum of \ps. However, X-ray bursts from LMXBs reach peak luminosities $L_x \gtrsim 10^{36} {\rm\ erg\ s}^{-1}$ which at the assumed distance is much greater than that of this source, and the timing characteristics are markedly different \citep{Strohmayer2006}. The light curve of \ps, Figure \ref{fig:ps_prop}, shows the source outburst reaches its peak after $\sim3,000$ s, and its flux seems to fade quickly ($>1,000$) after the maximum. In contrast to this behavior, the time profile of bursts from LMXBs usually display a steep rise, and exponential decay, and last for 10-100 s.  

The likely association between \ps\ and the stellar object 2MASS~J11500433--6224398 strongly suggests that the bursting source is the result of a stellar flare, since flaring stars also display temporal behavior similar to that observed for \ps,  and could account for the luminosity detected. However, the spectra from these sources is expected to be well described by a two-temperature plasma model, and the fit to the spectrum of \ps\ for such an emission profile yields a high-temperature component, $T_{\text{high}}\gtrsim 40$ keV, that is much hotter than observations suggest for stellar outbursts \citep[and references therein]{Feigelson1999}. This high temperature component is very significant, since it accounts for approximately $99\%$ of the $0.5-10.0$ keV X-ray flux. Further studies, in X-ray as well as in other bands, are needed to further constrain the nature of the source.

\section{CONCLUSIONS}

We present \xmm\ observations of the bright, nearby SNR \snr. Imaging and spectroscopy of this remnant are analyzed so as to characterize the nature of the emission in the X-ray band and determine the remnant evolutionary properties. Additionally, the characteristics of the transient source 2XMMi~J115004.8--622442, and its possible association with \snr, are studied. 

The X-ray emission from the SNR shows a spectrum best described by a thermal plasma model, with temperature $\sim 0.6$ keV. The absorbing column density to the SNR is low, \nH\ =2-4 $\times 10^{20} \text{cm}^{-2}$, which suggests the remnant is nearby. A distinct signature for excess nitrogen, and a deficit of oxygen (both relative to solar), are found to improve significantly the fits to the spectra of all regions of the SNR. These abundances are typical of wind material in RSG and Wolf-Rayet stars and suggest a massive progenitor for which the SNR emission is dominated by swept-up wind material. 

The detected X-ray source 2XMMi~J115004.8--622442, at the edge of the SNR, went into outburst during the observation period. It was undetected for the first 8 ks of the exposure, and displayed a significant increase in brightness in the X-ray band, reaching a peak count rate of 0.8 counts s$^{-1}$, for approximately 4,000 s. The spectral and temporal characteristics of 2XMMi~J115004.8--622442, together with its likely association with the object 2MASS~J11500433--6224398, are strongly indicative that this source is the result of a stellar flare.

\acknowledgments
The authors wish to thank Tea Temim, Jeremy Drake, Joseph Neilsen, Hans Moritz G\"unther, and Jeff McClintock for several helpful discussions. This work was supported in part by NASA grant NNX07AQ61G. P.O.S. acknowledges support from NASA contract NAS8-03060. B.M.G. acknowledges the support of a Federation Fellowship from the Australian Research Council through grant FF0561298. Financial support for XMM research by J.P.H. at Rutgers is provided by NASA grants NNX08AX55G, NNX08AX72G, and NNX09AAP40G. The MOST is operated with the support of the Australian Research Council and the School of Physics of the University of Sydney. This publication makes use of data products from the Two Micron All Sky Survey, which is a joint project of the University of Massachusetts and the Infrared Processing and Analysis Center/California Institute of Technology, funded by the National Aeronautics and Space Administration and the National Science Foundation.

\end{document}

%% file: tab1.tex
\begin{table*}
\begin{center}
\begin{threeparttable} 
\caption{\label{tab:obs} \xmm\ Observation Data}
\begin{tabular}{cccccccccccccc}
\toprule
\noalign{\smallskip}
\multicolumn{1}{c}{} & & \multicolumn{1}{c}{RA} & &
\multicolumn{1}{c}{Dec} & & \multicolumn{1}{c}{} & & \multicolumn{1}{c}{} & &
\multicolumn{3}{c}{Exposure (ks)}\\
\noalign{\smallskip}
\cline{11-13} 
\noalign{\smallskip}
\multicolumn{1}{c}{Obs. ID} & & \multicolumn{1}{c}{(J2000)} & & \multicolumn{1}{c}{(J2000)} & &
\multicolumn{1}{c}{Observation Date} & & \multicolumn{1}{c}{Instrument\tablenotemark{a}} & &
\multicolumn{1}{c}{Total} & & \multicolumn{1}{c}{Good\tablenotemark{b}} \\
\noalign{\smallskip}
\midrule
\noalign{\smallskip}                       
0503220101 & & 11 50 59.21 & & -62 20 26.4 & & 2007-07-07 & & PN  & & 
15.2 & & 6.4 
\\
(northwest, NW) & & & & & & & & MOS1  & & 19.3  & & 12.0
\\
 & & & & & & & & MOS2  & & 19.4 & & 12.1
\\
0503220201 & &11 50 17.92 & & -62 37 46.8 & & 2007-07-15 & & PN & & 
10.1 & & 0.0
\\
(southwest, SW) & & & & & & & & MOS1  & & 12.8 & & 2.3
\\
 & & & & & & & & MOS2  & &12.8 & & 2.2
\\
0503220301 & & 	11 53 19.39 & & -62 27 55.1 & & 2007-12-24 & & PN  & & 
12.3 & & 7.8
\\
(east, E) & & & & & & & & MOS1  & & 15.5 & & 7.7
\\
 & & & & & & & & MOS2  & & 15.5 & & 7.6
\\
\noalign{\smallskip}
\bottomrule
\noalign{\smallskip}

\end{tabular}

\begin{tablenotes}[para]\normalsize

\item[a]{All EPIC data was obtained in Full Frame mode, with the Medium filter.}

\item[b]{Total good time intervals remaining after flag and pattern filtering, and flare screening, for spectral analysis (\S 2).}

\end{tablenotes}
\end{threeparttable} 
\end{center}
\end{table*}

%% file: tab2.tex
\begin{table*}
\begin{center}
\begin{threeparttable} 
\caption[ ]{\label{tab:mosno2}
\xmm\ spectral results of regions using the VNEI plasma model. 
}
\begin{tabular}{cccccccccc} 
\toprule
\noalign{\smallskip}
 &  &$N_{\text{H}}$ & $kT$ & \multicolumn{4}{c}{Abundances (relative to solar)} & $n_{\text{e}}t$ &
$\chi^2$/ d.o.f 
\\
Region&Pointing &($10^{22}$ cm$^{-2}$) & (keV) &N & O & Ne & Mg& ($10^{11}$ s\,cm$^{-3}$) & 
\\
\noalign{\smallskip}
\midrule
\noalign{\smallskip}
a........... & NW & $0.02^{+0.01}_{-0.01} $& 0.63$^{+0.04}_{-0.04}$ & 
 1.85$^{+0.10}_{-0.10}$ &
 0.61$^{+0.02}_{-0.02}$ &
 1.21$^{+0.06}_{-0.05}$& 
 1.07$^{+0.09}_{-0.08}$&
 0.26$^{+0.29}_{-0.23}$ & 
 $685.0/247 = 2.77$
\\   
\noalign{\smallskip}
a1......... & NW & (0.02) \tablenotemark{a} & 0.78$^{+0.23}_{-0.15}$ & 
 2.13$^{+0.42}_{-0.35}$ &
 0.38$^{+0.05}_{-0.04}$ &
 0.80$^{+0.21}_{-0.19}$& 
 1.02$^{+0.31}_{-0.25}$&
 0.26$^{+0.10}_{-0.07}$ & 
 $134.0/151 = 0.89$
\\   
\noalign{\smallskip}
a2.........  & NW & (0.02) \tablenotemark{a} & 0.55$^{+0.05}_{-0.05}$ &
$2.00^{+0.29}_{-0.24}$ &
0.49$^{+0.03}_{-0.03}$&
1.17$^{+0.13}_{-0.12}$& 
1.15$^{+0.15}_{-0.14}$ & 
$0.47^{+0.12}_{-0.07}$ &
262.7/191 = 1.38
\\   
\noalign{\smallskip}
a3.........  & NW & (0.02) \tablenotemark{a} & 0.62$^{+0.05}_{-0.07}$ &
$1.67^{+0.15}_{-0.15}$ &
0.61$^{+0.03}_{-0.02}$&
1.15$^{+0.09}_{-0.08}$&
0.95$^{+0.14}_{-0.14}$ &
 0.26$^{+0.04}_{-0.04}$ &
313.9/203 = 1.55
\\   
\noalign{\smallskip}
a4.........  & NW & (0.02) \tablenotemark{a} & 0.54$^{+0.05}_{-0.07}$ &
1.88$^{+0.22}_{-0.20}$ &
0.75$^{+0.05}_{-0.04}$&
1.53$^{+0.13}_{-0.12}$& 
1.30$^{+0.26}_{-0.25}$ & 
0.25$^{+0.07}_{-0.04}$ &
199.7/180 = 1.11
\\   
\noalign{\smallskip}
a5.........  & NW & (0.02) \tablenotemark{a} & 0.56$^{+0.25}_{-0.17}$ &
2.01$^{+0.51}_{-0.45}$ &
0.79$^{+0.11}_{-0.10}$&
1.65$^{+0.34}_{-0.15}$& 
1.65$^{+0.79}_{-0.73}$ & 
0.23$^{+0.20}_{-0.09}$ &
114.7/131 = 0.88
\\   
\noalign{\smallskip}
b1.........  & SW & 0.01$^{+0.16}_{-0.01}$ & 0.63$^{+0.02}_{-0.06}$ &
1.69$^{+0.18}_{-0.11}$ &0.80$^{+0.04}_{-0.05}$&1.49$^{+0.11}_{-0.12}$&1.02$^{+0.25}_{-0.25}$ &0.23$^{+0.06}_{-0.04}$ &
149.5/92 = 1.62
\\   
\noalign{\smallskip}
b2.........  & SW & 0.04$^{+0.02}_{-0.02}$ &0.63$^{+0.11}_{-0.10}$ &
1.93$^{+0.28}_{-0.26}$ &0.81$^{+0.06}_{-0.06}$&1.75$^{+0.18}_{-0.17}$&1.31$^{+0.31}_{-0.28}$ & 0.21$^{+0.06}_{-0.04}$ &
118.9/93 = 1.28
\\   
\noalign{\smallskip}
c1.........  & E & 0.01$^{+0.12}_{-0.01}$ &0.75$^{+0.17}_{-0.13}$ &
1.49$^{+1.25}_{-0.83}$ &0.47$^{+0.26}_{-0.13}$&1.72$^{+0.75}_{-0.49}$& (1.00) \tablenotemark{b} & 0.13$^{+0.02}_{-0.02}$ &
32.1/37 = 0.89
\\   

\noalign{\smallskip}
c2.........  & E & 0.05$^{+0.13}_{-0.05}$ &0.73$^{+0.37}_{-0.27}$ &
1.29$^{+1.12}_{-0.75}$ &0.57$^{+0.25}_{-0.25}$&1.29$^{+1.13}_{-0.71}$& (1.00) \tablenotemark{b} & 0.11$^{+0.04}_{-0.05}$ &
46.1/40 = 1.15
\\   
 
\noalign{\smallskip}
\bottomrule
\noalign{\smallskip}
\noalign{\smallskip}
\end{tabular}

\begin{tablenotes}[para]\normalsize
\item[a]{All best-fit models to spectra from individual NW regions suggest absorbing column densities similar to that obtained by fitting the entire region (region a), and hence we adopt its value and fix it for the final individual fits.}

\item[b]{The fits to the spectra of regions c1 and c2, extracted from the faint eastern part of \snr\, are not significantly dependent on the relative abundance of Mg to solar levels, and this value was hence fixed at 1.00.}
\end{tablenotes}

\end{threeparttable} 
\end{center}
\end{table*}

%% file: tab3.tex
\begin{table}[h!]
\begin{center}
\begin{threeparttable} 
\caption{Observed Properties of \ps\ from \xmm\ data}
\label{tab:ps}
\begin{tabular}{lcc}
\toprule
Parameter&&Value \\
\midrule
Right Ascension, R.A. (J2000)&&$11^{\text{h}}50^{\text{m}}05^{\text{s}}$$\pm5''$\\
\noalign{\smallskip}
Declination, Dec. (J2000)&&$ -62^{\circ}24'43''$$\pm5''$\\
\noalign{\smallskip}
Column Density, $N_{\text{H}}$ ($10^{22} \text{cm}^{-2}$)&&0.07$^{+0.47}_{-0.05}$\\
\noalign{\smallskip}
Blackbody Temperature, $kT$ (keV)&&1.82$^{+0.27}_{-0.28}$\\
\noalign{\smallskip}
Spectral Index, $\Gamma$&&4.36$^{+2.46}_{-2.02}$\\
\noalign{\smallskip}
Total X-ray Flux \tablenotemark{a}, $F_X$ (erg cm$^{-2}$ s$^{-1}$)&&$3.02\times10^{-12}$\\
\noalign{\smallskip}
Blackbody Flux \tablenotemark{b}, $F_{\text{bb}}$ (erg cm$^{-2}$ s$^{-1}$)&&$2.80\times10^{-12}$\\
\noalign{\smallskip}
Reduced $\chi^2$ Statistic&&1.02(24 dof)\\
\noalign{\smallskip}
\bottomrule
\noalign{\smallskip}
\end{tabular}

\begin{tablenotes}[para]
\item[a]{$F_X$ is the average unabsorbed X-ray flux of \ps\ during the burst, calculated in the 0.3-10.0 keV energy band, using the best-fit blackbody + powerlaw model to the EPIC PN data.}
\item[a]{$F_{\text{bb}}$ is the X-ray flux from the blackbody component of the spectrum model of \ps\ during the burst, calculated in the 0.3-10.0 keV energy band.}
\end{tablenotes}

\end{threeparttable} 
\end{center}
\end{table}